\documentstyle[aps,aipbook,floats,epsf]{revtex}

\begin{document}
\title{Hydrodynamic Time Scales and Temporal Structure of GRBs }

\author{Re'em Sari$^*$ and Tsvi Piran$^{*,\dagger}$}  
\address{
$^*$Racah Institute for Physics, The Hebrew University, Jerusalem, Israel 91904 \\
$^\dagger$ITP, University of California, Santa Barbara, CA 93106, USA.}
\maketitle

\begin{abstract}
We calculate the hydrodynamic time scales for a spherical
ultra-relativistic shell that is decelerated by the ISM and discuss
the possible relations between these time scales and the observed
temporal structure in $\gamma$-ray bursts. We suggest that the bursts'
duration is related to the deceleration time, the variability is
related to the ISM inhomogeneities and precursors are related to
internal shocks within the shell. Good agreement can be achieved for
these quantities with reasonable, not fined tuned, astrophysical
parameters.  The difference between Newtonian and relativistic reverse
shocks may lead to the observed bimodal distribution of bursts'
durations.
\end{abstract}

\section*{Introduction}

Gamma-ray bursts (GRBs) are most likely generated during deceleration
of ultra-relativistic particles.  A cosmological compact source that
emits the energy required for a GRB cannot generate the observed non
thermal burst.  Instead it will create an opaque fireball (see e.g.
\cite{piran94,piran95}).  If even a small amount of baryonic matter is
present then the ultimate result of this fireball will be a shell of
ultra-relativistic particles \cite{shemipir}.  The kinetic energy can
be recovered as radiation only if these particles are decelerated by
the ISM \cite{mesrees92} or by internal shocks \cite{reesmes94,narpacpir}.
We show that a careful analysis of the interaction between the
ultra-relativistic particles and the ISM  changes some of the previous
results and sheds a new light on the expected temporal structure in GRBs.

\section*{Planar Symmetry}

Consider a slab of ultra-relativistic cold dense matter
with a Lorentz factor $\gamma \gg 1$ that hits a stationary cold
interstellar medium - the ISM. Two shocks form: a reverse
shock that propagates into the dense relativistic shell, reducing its
speed and increasing its internal energy, and a forward shock that
propagates into the ISM giving it relativistic velocities and internal
energy. A contact discontinuity separates the shocked shell material
 and the shocked ISM. 

There are two  limits  \cite{sarpira} in which the reverse shock is
either Newtonian or ultra-relativistic (the forward shock is always
ultra-relativistic in our case). We define $f$ as the density ratio of the
ultra-relativistic shell and the ISM .
If $\gamma^2\gg f$ the reverse shock is ultra-relativistic.  In this
case:
\begin{equation}
\label{eq:rel:g3bar}\gamma _2=\gamma ^{1/2}f^{1/4}/\sqrt{2} 
\ \ ;\ \ t_\Delta=\Delta \gamma \sqrt{f}/2c \ ,
\end{equation}
where $\gamma_2$ is the Lorentz factors of the shocked
material (shocked material on both sides of the contact 
discontinuity move at the same velocity) relative to an observer at 
infinity, $\Delta$ is the width of
the ultra-relativistic shell and $t_\Delta$ is the time that the
reverse shock crosses the shell.  Since $\gamma _2\ll \gamma $, almost
all of the initial kinetic energy is converted by the shocks into
internal energy. Therefore the process is over after a single passage
of the reverse shock through the shell, and the relevant time scale
for energy extraction is the shell crossing time, $t_\Delta$. Along
the contact discontinuity the energy densities are equal and since both
shocked regions have comparable width they release comparable amounts
of energy.  The ISM mass swept by the forward shock at the time that
the reverse shock crosses the shell is $\sim f^{-1/2}$ of the shell's
mass.  This is larger than the simple estimate given by M\'es\'zaros,
\& Rees \cite{mesrees92} of $\sim \gamma ^{-1}$. 

If $f \gg \gamma^2$ the reverse shock is Newtonian  
and: 
\begin{equation}
\label{eq:g2:newtonian}\gamma _2 \cong \gamma 
\ \ ; \ \ t_\Delta=\sqrt{9/14}\Delta \gamma \sqrt{f} /c
\end{equation}
Since now $\gamma _2 \cong \gamma$ the reverse shock converts only a small 
fraction, $\gamma /\sqrt{f}\ll 1$, of
the kinetic energy into internal energy and $t_\Delta$ is no longer the
relevant time scale for energy extraction. 
The main deceleration is during 
a quasi-steady state in which the shell decelerates continuously 
without shocks.
The slowing down time can be estimated 
by $\sim \Delta f/c \gamma $. During this time
the forward shock collects a fraction $\sim \gamma ^{-1}$ of the
shell's rest mass, which is the same as the original estimate of
M\'es\'zaros \& Rees \cite{mesrees92}. 
In contrary to the relativistic case,
there are two time scales now: the shock crossing
time, $t_\Delta$, and the total slowing down time, $t_\gamma$.

In the realistic situation the ISM density is probably inhomogeneous.
Consider a density jump by a factor $f^{\prime }$ over a distance
$l_{ISM}$.  The forward shock propagates into the ISM with a density
$n_1$ as before and when it reaches the position where the ISM density
is $ n_1 f^{\prime }$ a new shock wave is reflected.
This shock is reflected
again of the shell. Similar analysis shows that the reflections time
is $\sim l_{ISM}/4c \sqrt{f^{\prime }}$.

Finally, we mention the possibility of internal shocks inside the
shell \cite{reesmes94}.
These may form when faster material overtakes slower material.
If the Lorentz factor varies by a factor of $\sim 2$ over a length
scale $\delta R\leq \Delta $ then the time for these shock to from is
$\sim \delta R\gamma ^2 /c<\Delta \gamma ^2/c$.  This time scale is
shorter than the slowing-down time scale and therefore internal shocks
appear before considerable deceleration in the Newtonian case. In the
relativistic case considerable deceleration occurs before internal
shocks unless $\delta R \ll \Delta$.

\section*{Spherical Considerations}

In a spherical system 
the density ratio $f\sim R^{-2}$ decreases with
time.  Initially $f/\gamma^2\gg 1$ and the reverse shock is Newtonian.
The energy conversion depends critically on the question whether this
shock become relativistic before the kinetic energy is extracted from
the shell. This depends, in turn, 
on the ratio of two radii: 
$R_N= l^{3/2}/\Delta^{1/2} \gamma^2 $ where
$f/\gamma^2=1$ and the reverse shock becomes relativistic and
$R_\Delta=l^{3/4} \Delta^{1/4}$ where the reserve shock crosses the shell.
The radius   
$l \equiv (E /n_1 m_p c^2)^{1/3}$ in these expressions is the Sedov length
which is familiar from SNR theory. 
Two other
important radii are: $R_\gamma=l/\gamma^{2/3}$ where the forward shock 
sweeps a mass
$M/\gamma$ ($M$ is the shell's rest mass) and $R_s=\Delta \gamma^2$ 
where the shell
begins to spread if the initial Lorentz factor varies by order
$\gamma$ 
\cite{piran94}. Note that $R_s$ is also an upper limit
for the the location of internal shocks since $\delta R < \Delta$.
Conveniently, the
four critical radii are related by one dimensionless quantity:

\begin{equation}
\label{eq:xi}\xi \equiv { {\left({l/\Delta}\right)}^{1/2} \gamma^{-4/3}}
\ \ \ ; \ \ \ R_N/\xi = R_\gamma = \sqrt \xi R_\Delta = \xi^2 R_s \ .
\end{equation}

Two possibilities exist:

1. $\xi >1$ - the Newtonian case: $R_s<R_\Delta<R_\gamma<R_N$ and
shock reaches the inner edge of the shell while it is still Newtonian.
Most of the energy is extracted during a steady state deceleration
phase within the radius $R_\gamma$.  Since $R_s$ is smaller than
all other radii spreading might be important. If the shell is
spreading then $\Delta$ in the above expressions should be replaced by
$R/\gamma^2$. This delays the time at which the reverse shock reaches
the shell and decreases the shell's density. These effects lead to a
triple coincidence: $R_\Delta = R_\gamma=R_N$ with $\xi\approx 1$ and
a mildly relativistic reverse shock during the period of effective
energy extraction.  Without spreading only a small fraction of the
total energy is converted to thermal energy in the reverse shock. With
spreading both shocks convert comparable amounts of energy.

2. $\xi <1$ - the relativistic case: $R_N<R\gamma <R_\Delta < R_s$.
The reverse shock becomes relativistic before it crosses the shell.
Only a small fraction of the energy is converted at $R_\gamma $ and
the kinetic energy is converted into internal energy only at $R_\Delta
$. $R_s$ is larger than all other radii and spreading is unimportant.
It is interesting to note that in this limit $\gamma _2(R_\Delta )\sim
(l/ \Delta)^{3/8}$ is independent of the initial Lorentz factor
$\gamma $ and it is only weakly dependent on other parameters.  This
might have an important role in the fact that the observed radiation
always appears as low energy $\gamma $-rays.

Neither the internal shocks nor the ISM inhomogeneity time scales are
affected by these spherical considerations. The former depends only
upon $\delta R$, $\Delta$ and $\gamma$ and the latter depends only on
$l_{ISM}$ and $\gamma$.  Both are constant throughout the spherical
expansions.

\section*{Observational Implications to GRB}

We examine now the
possible relation between the observed time scales and the
hydrodynamic time scales. We assume that the shocked material 
emits the radiation on a time scale shorter than the
hydrodynamic time scales. A simple estimate of synchrotron cooling
rate (assuming equipartition of the magnetic field energy) is
consistent with this assumption. The
requirement that the cooling time is shorter than the observed
variability imposes interesting constraints on the physical conditions
within the shocks.  In particular it demands that the turbulent
magnetic field within the shocked material should be very close to
equipartition with the thermal energy there. We discuss these
conditions elsewhere \cite{narpirsar}.

While the value of $l \approx 10^{18}$cm is known
(Using $E=10^{51}$ergs and $n_1=1$particle/cm$^3$),
the values of $\Delta $ and $\gamma $ are more
ambiguous. Using the canonical values \cite{piran95} $\gamma =10^3$
and $\Delta =10^7$cm we get $\xi \cong 30>1$, corresponding to a
Newtonian reverse shock.  Nevertheless a value of $\xi \cong 0.1<1$ is
also possible with reasonable parameters (for example $\Delta =10^9$cm
and $\gamma =10^4$). Therefore, both relativistic and Newtonian reverse
shock are possible.

\begin{figure}
\begin{center}
\leavevmode
\epsfxsize=150pt \epsfbox{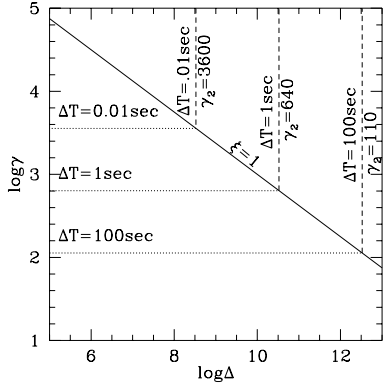}
\end{center}
\caption{
The observed duration of the burst (dotted curve in the
Newtonian regime and dashed curve in the relativistic regime) as
function of $\gamma $ and $\Delta $ for $l=10^{18}$cm. 
The thick line ($\xi=1$) separates the
Newtonian (lower left) and the Relativistic (upper right) regions.
Spreading drives all Newtonian cases to the $\xi\approx 1$
line.
}
\end{figure}

The bursts' duration is determined by the slowing down time of the
shell divided by $\gamma_2^2$.
Thus, given a typical radius of
energy conversion, $R_e$ the observed time scale is:
\begin{equation}
\label{tobs}
\Delta t_{obs}=R_e/\gamma_2^2 c=
\cases { \Delta /c &  if $\xi<1$ ~~  (Relativistic); \cr 
R_\gamma /\gamma ^2c\sim l /\gamma^{8/3}  c         & 
if $\xi > 1 $ ~~  (Newtonian) \ . \cr }
\end{equation}
$\Delta t_{obs}$ ranges from $\sim 1 $msec, for $\Delta = 3 \times
10^7$cm and $\gamma =10^4$, to $\sim 100$sec for $\gamma =10^2$ and
$\Delta = 10^{13}$cm (see fig. 1).  
The observed
durations of the brightest 30 bursts limit $\gamma $ to $100<\gamma
<10^4$ with a typical value of $\approx 500$ and it limits $\Delta$ to
$\Delta < 3 \times 10^{12}$cm.

A possible source of the observed fluctuations
during the bursts is inhomogeneity in
the ISM.  If the length scale of the inhomogeneity is $l_{ISM}$ and
the density varies by one order of magnitude then the time scale for
the observed variability will be:
\begin{equation}
\label{tvarISM}t_{var}\sim l_{ISM}/ 10 \gamma_2^2 c \ .
\end{equation}
Clearly, $t_{var}$  can be sufficiently short 
if $l_{ISM}$ is sufficiently small.

Precursors which appear in about 3\% of the bursts might be explained by
internal shocks that take place at $R \approx \delta R \gamma^2 \le
\Delta \gamma^2$ while
the main burst originates from the interaction with the ISM.
The duration of the precursor is
$\Delta t_{pre}=\Delta /c$, which requires values of $\Delta $ as high
as $10^{10}-10^{12}$cm to produce the observed precursors of
$1-100\sec $.  If $\xi>1$ the main bursts will have a duration $\Delta
t_{obs} \approx l/\gamma^{8/3} c = \xi^2 \Delta$.  This will also be
the typical separation $\Delta t_{pre-maim}$ between the precursor and
the main burst. We expect a time delay between the precursor and the
main burst which will be comparable to the duration of the main
burst. Note that a correlation of the form $\Delta t_{pre-maim}\approx
4.5\Delta t_{obs}$ exists (but was not reported) in the data of Koshut
{\it et. al., } \cite{koshut}. If $\xi<1$ then 
precursors do not occur (unless $\delta R \ll \Delta$).
This is in agreement with the lack of observed
precursors in short bursts.

\section*{Conclusions}

We have calculated the hydrodynamic time scales of shocks during the
interaction between an ultra-relativistic shell and the ISM.
We find that with reasonable
astrophysical parameters these time scales are in a good agreement
with the observed time scales in GRBs. Our analysis shows that there
are two kinds of shocks: Newtonian and Relativistic. The difference
between them might correspond to the observed bimodality of
short and long bursts. Finally, we
suggest that precursors might be emitted due to internal shocks within
the ultra-relativistic shell while the main burst emerges later from
the interaction with the ISM.

This research was supported by BRF grant to the
Hebrew University by NASA grant NAG5-1904 to Harvard University
and by NSF grant PHY94-07194 to 
the ITP.

\end{document}